\documentclass[runningheads]{llncs}

\usepackage{cite}
\usepackage{xcolor}
\usepackage{tabularx}
\usepackage[hidelinks]{hyperref}
\usepackage{orcidlink}
\usepackage[htt]{hyphenat}

\usepackage[caption=false]{subfig}

\usepackage[T1]{fontenc}
\usepackage{graphicx}

\usepackage{url}

\newif\ifdraft
\drafttrue
\ifdraft
\newcommand{\spnote}[1]{ {\textcolor{magenta} { ***SP: #1 }}}
\newcommand{\bsnote}[1]{ {\textcolor{red} { ***BS: #1 }}}
\newcommand{\mlnote}[1]{ {\textcolor{blue} { ***ML: #1 }}}
\newcommand{\red}[1]{ {\textcolor{red} {\em #1 }}}
\else
\newcommand{\spnote}[1]{}
\newcommand{\bsnote}[1]{}
\newcommand{\mlnote}[1]{}
\newcommand{\red}[1]{}
\fi

\newcommand{\code}[1]{\texttt{#1}}

\begin{document}

\title{A fast MPI-based Distributed Hash-Table as Surrogate Model demonstrated in a coupled reactive transport HPC simulation}
\titlerunning{MPI-based DHT for HPC}

\author{Max L\"ubke\inst{1}\orcidlink{0009-0008-9773-3038} \and
  Marco {De Lucia}\inst{2}\orcidlink{0000-0002-1186-4491} \and
  Steffen Christgau\inst{3}\orcidlink{0000-0002-8702-8422} \and
  Stefan Petri\inst{4}\orcidlink{0000-0002-4379-4643} \and
  Bettina Schnor\inst{1}\orcidlink{0000-0001-7369-8057}}

\institute{University of Potsdam, Institute of Computer Science, An der Bahn 2,
  14476 Potsdam, Germany\\
  \email{\{mluebke,schnor\}@uni-potsdam.de}\and
  GFZ Helmholtz Centre for Geosciences, Telegrafenberg, 14473 Potsdam, Germany\\
  \email{delucia@gfz.de}\and
  Zuse Institute Berlin, Takustr. 7, 14195 Berlin, Germany\\
  \email{christgau@zib.de} \and
  Potsdam Institute for Climate Impact Research, Member of the Leibniz Association, Telegrafenberg, 14473 Potsdam, Germany\\
  \email{petri@pik-potsdam.de}
}

\maketitle

\begin{abstract}   

This is an extended version of the paper "A fast MPI-based Distributed
  Hash-Table as Surrogate Model for HPC Applications", which appeared in the
  Proceedings of the 25th International Conference on Computational Science
  (ICCS)~\cite{LLPS25b}. This extended version provides more details about the
  approaches, implementation, and evaluation of the MPI-DHT. Additionally, it
  includes a comparison of the MPI-DHT with the DAOS key-value
store.

  Surrogate models can play a pivotal role in
  enhancing performance in contemporary High-Performance
  Computing applications. Cache-based surrogates
  use already calculated simulation results to
  interpolate or extrapolate further simulation output
  values.  But this approach only pays off if the access
  time to retrieve the needed values is much faster than
  the actual simulation.
  While the most existing key-value stores use a Client-Server
  architecture with dedicated storage nodes, this is not the most
  suitable architecture for HPC applications. Instead, we
  propose a distributed architecture where the parallel processes
  offer a part of their available memory to build a shared
  distributed hash table based on MPI.
  This paper presents three DHT approaches with the special
  requirements of HPC applications in mind.
  The presented lock-free design outperforms both DHT
  versions which use explicit synchronization by coarse-grained
  resp.\ fine-grained locking. The lock-free DHT shows very good
  scaling regarding read and write performance. The runtime
  of a coupled reactive transport simulation was improved between 14\% and
  42\% using the lock-free DHT as a surrogate model.

  \keywords{Distributed Hash Table \and Key-Value Store \and Surrogate Model \and RDMA \and DAOS}
\end{abstract}

\section{Introduction}\label{sec:introduction}

During the last years, several research groups have shown the benefit of
surrogate models for accelerating parallel
simulations~\cite{2019-Raissi,Reaktoro2020,LKLLS21, deluciaDecTreeV10Chemistry2021}.  While Raissi et
al.~\cite{2019-Raissi} propose {\em physics-informed neural networks}, De Lucia et al.\ use
  {\em decision trees} to speed up geochemical simulations~\cite{deluciaDecTreeV10Chemistry2021}.
Reaktoro~\cite{Reaktoro2020} and POET~\cite{LKLLS21} make use of fast caches
within their surrogate approach.  Reaktoro uses so called {\em On-Demand Machine
    Learning} (ODML) to speed up chemical calculations.  ODML works by extrapolating
new chemical states of multiphase systems undergoing chemical reactions.  Given
new inputs, Reaktoro's ODML looks up a past calculation for which the Jacobian
is known and estimates the output using Taylor
extrapolation~\cite{Reaktoro2020}. POET is a coupled reactive transport
simulation which uses a distributed hash table (DHT) as fast storage for
simulation results.  These are reused to approximate further results in
subsequent time steps.  An {\em approximated lookup} is used where the modeler
has to specify a possible rounding for each variable.  Thus, there is a
tradeoff between performance and modeling accuracy.
Such cache-based surrogates work fine as long as the query and
retrieval times are much faster than the full physics geochemical
simulations.

Starting with the
relevance of key-value stores in the context of big social media applications,
see for example Facebook~\cite{2013-Facebook}, researchers spent great effort to
improve the performance.
One of the first steps was to accelerate the communication
by introducing RDMA-capable networks~\cite{2011-MemcachedPanda,2013-Pilaf,2014-FaRM,2014-HERD,NESSIE2017,DBLP:journals/sp/GerstenbergerBH14,UPC-OSC2011,LKLLS21}.
All these RDMA-capable approaches have in common that they have to
solve three problems:

\begin{enumerate}
  \item Design of {\bf Addressing}: For an RDMA operation, the source has to know the
        physical address of the object where to read resp.\ write the data on
        the target side. Some systems use a special {\em index table} which
        stores the mapping between key and address~\cite{NESSIE2017, 2021-RACE}.
        Others use a {\em direct hashing} where the physical address is directly
        derived from the key~\cite{LKLLS21}.  In case of a client-server
        architecture, the address information is exchanged via a Remote
        Procedure Call (RPC) in
        advance~\cite{2014-HERD,2015-MDHIM,2020-DAOS}.
  \item Design of {\bf Data Consistency}: Since read and write operations may occur
        concurrently, there must be taken care of data consistency. This is done
        by a server (central
        approach)~\cite{2014-FaRM,2014-HERD,2015-MDHIM,2020-DAOS}, a
        synchronization protocol (for example Readers \&
        Writers)~\cite{LKLLS21}, or using checksums for self-verifying data
        structures~\cite{2013-Pilaf,2021-RACE}.
  \item Design of {\bf Collision Handling}: There exist
        several collision handling strategies. For example, Hopscotch hashing
        implemented by FaRM~\cite{2014-FaRM} or Cuckoo hashing implemented by
        Pilaf~\cite{2013-Pilaf} and LFBCH~\cite{2023-EuroPar-LFBCH}.
\end{enumerate}

MPI is the de facto standard in HPC and also offers an API for Remote Memory
Access (RMA).  Hence, it is a straightforward idea to implement the key-value
store using the MPI RMA API and to benefit easily from the experiences and
improvements of the HPC community (see for
example~\cite{DBLP:journals/sp/GerstenbergerBH14,UPC-OSC2011,LKLLS21}).
With every optimization in the MPI library, DHTs can profit from it
without the necessity of reimplementation in a self-made communication layer.
In an earlier work, we integrated an MPI-based DHT, called MPI-DHT in
the following, into the coupled reactive transport simulator
POET as a surrogate model~\cite{LKLLS21}.

\begin{figure}
  \centering
  \subfloat[Fully Distributed]{%
    \hspace{1cm}
    \includegraphics[height=.2\textheight,clip]{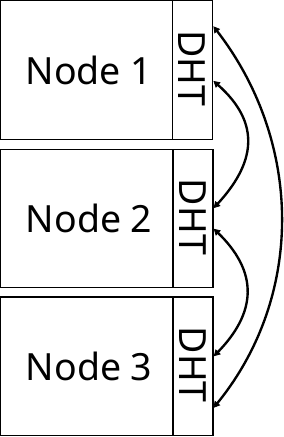}%
    \hspace{1cm}
  }
  \subfloat[Server-based]{%
    \includegraphics[height=.2\textheight,clip]{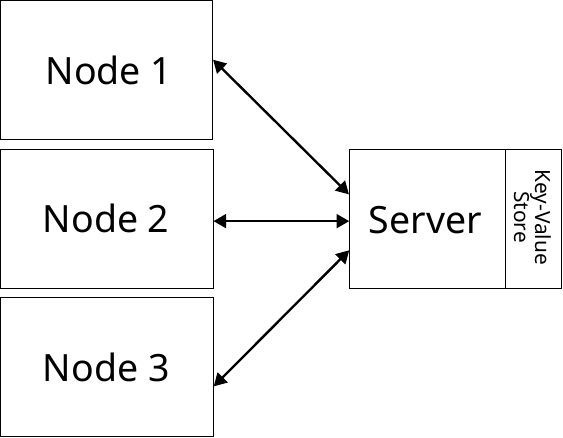}%
  }
  \caption{Comparison of distributed and server-based key-value stores.}
  \label{fig:comparison_DHT}
\end{figure}

Most key-value stores are server-based, see for
example~\cite{2014-FaRM,2013-Pilaf,2014-HERD,2011-MemcachedPanda,2020-DAOS}.
However, this architecture is not the most suitable for HPC applications
because it requires additional hardware and setup as shown in
Figure~\ref{fig:comparison_DHT}. Further, Figure~\ref{fig:comparison_DHT} shows
the different communication patterns of a fully distributed and a server-based
key-value store. While in the server-based approach, the clients always
communicate with a central server, in the fully distributed approach clients
may access the data directly from remote storage, e.g.\ using RDMA operations.
The benefit of the central server approach is data consistency, while in the
distributed approach data consistency can only be achieved by additional
synchronization protocols. This paper discusses fully distributed key-value
store architectures based on MPI that allow seamless integration into
scientific applications.

The contributions of this work are:
\begin{itemize}

  \item A performance comparison of the MPI-based DHT (MPI-DHT)
    from~\cite{LKLLS21} and the server-based key-value store
    DAOS~\cite{2020-DAOS}. Despite a scaling problem, the distributed
    architecture of the MPI-DHT performs much better than the server-based DAOS
    architecture.
  \item Presenting two additional distributed hash-table architectures based on
        the MPI one-sided communication API, including a fine-grained locking DHT, and
      one lock-free approach which uses optimistic concurrency control.
  \item An evaluation of all three DHT approaches with a synthetic
        benchmark testing the DHT read and write performance shows the
        performance advantages of the lock-free
        approach by a factor of up to 1,400.
  \item The lock-free DHT scales very well, both for read and
        write requests. For 640~participating processes, 16 Million
        read operations per second and 15 Million
        write operations per second
        were observed.
  \item The DHT was integrated as fast data cache into a coupled
        geochemical transport simulation. In case of the lock-free DHT, the
        runtime was improved by~42\% running on 128~cores, and by 14\%
        running on 640~cores.

\end{itemize}

\section{Related Work}

Since more than one decade, researchers  investigate implementation
possibilities for key-value store
and DHTs which exploit the capabilities of modern networks.

Gerstenberger et al.\ presented a distributed hash table using MPI RMA  for the Cray Gemini and Aries
interconnects~\cite{DBLP:journals/sp/GerstenbergerBH14}.
For the coordination of reader and writer processes, they use
MPI's passive target synchronization with a best effort approach: If a
lock acquisition does not succeed, all data structures will be
released and the process will try again later with an exponential back
off.  This is implemented within their own MPI-3.0\,RMA library
implementation optimized for the Cray Gemini and Aries interconnects
where inserts are based on atomic compare and swap (CAS) and atomic
\code{fetch\_and\_op} operations which are implemented on top of
proprietary Cray-specific APIs. The measurements demonstrate the
scaling of the implementation on up to 32k\,cores on the Cray machine.
Another MPI-based distributed hash table is the MDHIM key-value
store~\cite{2015-MDHIM}. The evaluation shows that it outperforms
the Cassandra key-value store which is common in the field of Big Data
using a uniform distribution. MDHIM does not exploit the RDMA capability
of the underlying network. Instead, each MPI process starts a so-called
Range Server which serve  client requests.

There exist several key-value stores leveraging RDMA.
Pilaf~\cite{2013-Pilaf}, FaRM~\cite{2014-FaRM}, and HERD~\cite{2014-HERD}
have a server-based architecture which is not suited for HPC
environments.
Instead, we investigate
different DHT implementations which have a fully distributed architecture.
The MPI-DHT versions which are discussed in this paper
implement the read operation similar to the Cuckoo hashing approach
from Pilaf, as we provide multiple indices where a value might be stored.
However, our implementation does not
require any movement of buckets during a write process, as seen in Cuckoo
hashing~\cite{2013-Pilaf, NESSIE2017} or hopscotch hashing~\cite{2014-FaRM}.

Nessie also uses
RDMA read and write operations, but it has a distributed architecture.
Further, Nessie is
designed to efficiently support large data values up to several kilobytes~\cite{NESSIE2017}.
Therefore, Nessie decouples indexing metadata, so-called index
tables, and the key-value pairs.
To ensure consistency, Nessie uses atomic swap
operations on the index table.

In the following we present different MPI-based DHT implementations
which all use RDMA for read and write accesses.

\section{Comparison of MPI-DHT and the Key-Value Store DAOS}\label{sec:comparison}

Available implementations for production use in HPC are rare. An exception are
the MPI-DHT, and the  DAOS
implementation\footnote{\url{https://github.com/daos-stack/daos}}. DAOS stands
for Distributed Asynchronous Object Storage and is an open-source server-based
object store designed by Intel for distributed Non-Volatile Memory (NVM)
technologies~\cite{2020-DAOS}. We conducted a comparative evaluation between the
MPI-DHT~\cite{LKLLS21} and the server-based approach DAOS. 

Next we describe the key design features regarding addressing, data consistency
and collision handling of the two systems.

\subsection{Coarse-Grained Lock-Based MPI-DHT}\label{secsec:coarseDHT}

The MPI-DHT API~\cite{LKLLS21} consists of only four operations:
\texttt{DHT\_create, DHT\_read, DHT\_write} and
\texttt{DHT\_free}. The \texttt{DHT\_read} and \texttt{DHT\_write}
operations are implemented using MPI's one-sided communication
operations \texttt{MPI\_Put} resp.\ \texttt{MPI\_Get} to access the
remote memory.

The original MPI-DHT implementation utilizes MPI's passive target
synchronization, leveraging \texttt{MPI\_Win\_lock} and
\texttt{MPI\_Win\_unlock} operations~\cite{LKLLS21}. Each participating process
allocates memory using \texttt{MPI\_Window} to share it with the other
application processes. The memory window is subsequently divided into multiple
buckets. Each bucket contains a key-value pair and an additional Byte for meta
information, such as an {\em occupied} flag. Thus, this approach requires only
one additional Byte per bucket.

  {\bf Addressing:} The address of a bucket is a pair
consisting of the process rank (target rank) holding the data and an
index within the memory window.  First, a 64-bit hash sum of the key
is generated. A modulo operation by the total count of participating
processes is used to determine the target rank.

  {\bf Collision Handling:} A set of bucket indices is derived from the
hash sum by dividing the hash into smaller parts. Given an index of
size $n$ bytes and $B$ buckets per memory window, we choose $n$ as the
smallest integer that satisfies the inequality $log_2(B) \leq 8n$.

\begin{figure}
  \centering
  \includegraphics[width=0.6\textwidth,clip]{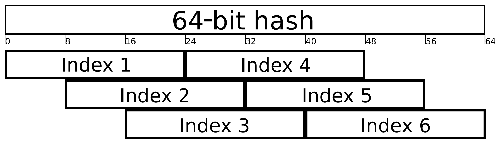}
  \caption{Determination of addresses of a bucket in the MPI-DHT.~\cite{LKLLS21}.}
  \label{fig:DHT_addressing}
\end{figure}

Figure~\ref{fig:DHT_addressing} shows an example where a 3-byte index is used
which allows a memory region with a maximum of $2^{8 \cdot 3} = 2^{24}$
buckets. By moving forward by 1 Byte, 6 different bucket indices are derived at
which a key-value store can be stored in the memory window of the target rank.

In the context of a write operation, the target rank and the corresponding
bucket indices are calculated, and then the first bucket is checked using
\texttt{MPI\_Get}. If the meta information is not set to {\em occupied}, data
are written to the bucket using \texttt{MPI\_Put}.  The data consist of the
key-value pair and the meta information where the {\em occupied} flag is set.
In the event that the bucket is already {\em occupied}, but the requested key
matches the current key in the bucket, the value is updated. If the key does
not match, the bucket at  the next index is checked. This process is repeated
until an empty bucket or a bucket with the same key is found. If the last index
is reached and no empty bucket or updatable key was found, this bucket is
overwritten.

The read operation employs a similar scheme, traversing all possible
buckets of the target rank. It is checked, whether a bucket is {\em
    occupied} and whether the key matches the requested key. If a match exists,
the value is returned.

  {\bf Data Consistency:} The MPI-DHT implements a Readers\&Writers
semantic.  Whenever an \texttt{DHT\_read} or \texttt{DHT\_write}
operation is initiated on a target rank, the entire memory window of
the target rank is locked with \texttt{MPI\_Win\_lock}. Depending on
whether the operation is a \texttt{DHT\_read} or \texttt{DHT\_write}
operation, the window is locked in {\em shared} or {\em exclusive} mode. The lock
is released with \texttt{MPI\_Win\_unlock} after the operation is
completed.  Since always the entire memory window is locked, we call
it the coarse-grained MPI-DHT.

\subsection{DAOS}

The DAOS server is responsible for determining memory addresses,
ensuring the consistency of data, and handling collisions.
The protocol initiates with a request message to the server. In the
case of a read operation, the server is informed of the location on
the client side where the data is to be written and initiates an RDMA
\texttt{PUT} operation.  In the case of a write operation, the message
includes the address on the client side where the data is to be
retrieved. The server then retrieves the key-value pair to be stored
by a subsequent call to a RDMA \texttt{GET} operation. After all RMA
operations have been completed, the server sends a reply message to
inform the client that the requested operation is finished.

The DAOS protocol differentiates between small and large
messages. For messages smaller than 18 kilobytes, the initial request
message already contains the data to be written or read, thus
eliminating the need for additional RMA operations on the server side.

\subsection{Test bed and Benchmark Description}
\label{sec:daos_testbed}

We conducted the evaluation on four nodes of the Turing Cluster at the
University of Potsdam.
Each node is equipped with two Intel Xeon E5-26540v4 processors with 12
hardware cores each, 64 GBytes of main memory, and interconnected through
RoCE ConnectX-6 Dx 100 Gbit NICs.

The nodes were configured with Rocky Linux 8.8 and Mellanox OFED driver version
1.0.1. One node hosted the DAOS server, while the remaining three nodes served
as clients.
Since the MPI-DHT stores the values into the RAM, DAOS was also configured to
use non-persistent RAM as key-value storage. For the MPI-DHT measurements,
Open MPI version 4.1.5a1 was used.

We measure the operations per second (OPS) for \texttt{Read} and
\texttt{Write} operations separately across varying numbers of
clients. For the DAOS evaluation, we utilized the KV-API of the
library to interact with the server.  The key and value sizes were set
to 80 bytes and 104 bytes, respectively, in accordance with the
requirements of the chemical simulation described in the POET
evaluation~\cite{LKLLS21}, which is greater than the 16 byte key and
32 byte value sizes used for the evaluation of other key-value
stores~\cite{2014-HERD,2014-FaRM,2021-RACE}. Keys were randomly
generated using a uniform distribution where every client starts with
a different seed. Each client executes 100,000 write operations and
then reads all 100,000 written entries.  The median read and write
performance of five repetitions are presented without error bars, as
the maximum coefficient of variation is 3.8\%.

The benchmark was scaled from 12 to 72 clients, with an increase of 12
clients per step. The processes were mapped onto a node in a dense
manner, resulting in the filling of one NUMA node after another. For
the distributed MPI-DHT, each client contributed 1 GByte of its memory
to the DHT.

\subsection{Results: Read/Write Performance}
\label{sec:results_daos_mpi}

\begin{figure}
  \centering
  \includegraphics[width=0.65\textwidth,clip]{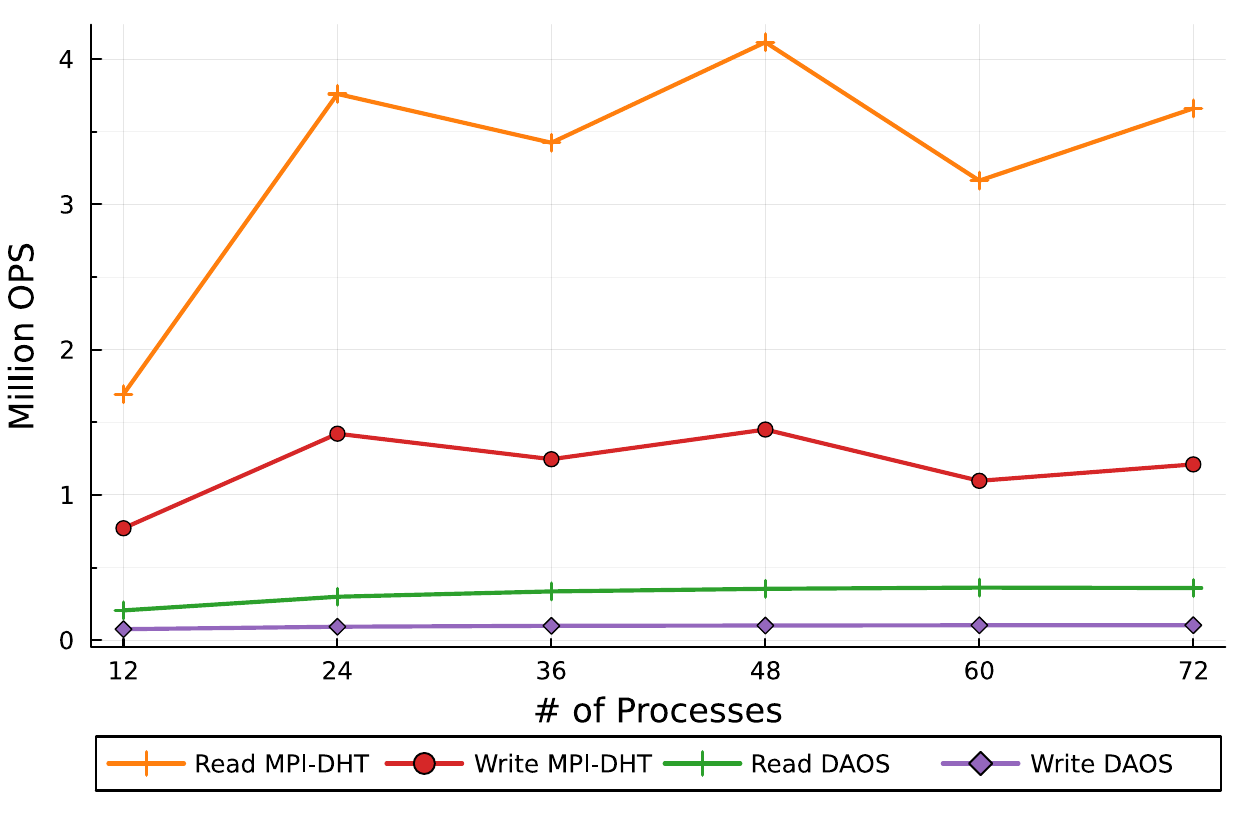}
  \caption{Throughput of DAOS and MPI-DHT for \textit{read} and \textit{write} operations.}
  \label{fig:DAOS_OPS}
\end{figure}

Obviously, the MPI-DHT outperforms DAOS in both read and write
operations, as illustrated in Figure~\ref{fig:DAOS_OPS}.  The MPI-DHT
reaches its peak performance with 4.12 million reads and 1.45 million
writes per second when utilizing two nodes (48 processes).  The
throughput of DAOS remains relatively constant and reaches its maximum
read rate at 60 clients with 362~thousand operations per
second and its maximum write rate with 103~thousand operations per second
at 72 processes.

The median latency of the MPI-DHT is
between $4\,\mu s$ and $17\,\mu s$ for read access and between
$13\,\mu s$ and $57\,\mu s$ for write access. The latency of DAOS is
between $56\,\mu s$ and $198\,\mu s$ for read and between $157\,\mu
  s$ and $698\,\mu s$ for write access. The factor of improvement of
the MPI-DHT compared to DAOS lies between 8.2 - 12.5 for read and
10.1 - 15.3 for write operations.

Instead of scaling with the number of clients,
MPI-DHT and DAOS demonstrate both a stagnation in throughput.
The poorer DAOS performance can be attributed to the central approach and
indicates the server as the bottleneck of the setup. The performance of the
MPI-DHT is deeper analyzed in the following.

\subsection{Discussion of MPI-DHT Performance}\label{sec:mpi-dht-discussion}

We identified the synchronization overhead
caused by the explicit locking of a target window as specified by the MPI
standard as the bottleneck. To be concise, for 72 processes 80~\%
(\texttt{DHT\_write}) resp.\ 48~\% (\texttt{DHT\_read}) of one call are spent in
\texttt{MPI\_Win\_Lock} and \texttt{MPI\_Win\_Unlock}.
There are two reasons for this high synchronization overhead. Firstly,
the MPI-DHT treats the entire DHT memory of one process as a single window,
which results in the unnecessary locking of the entire memory while only one
bucket is accessed.
Secondly, in the Open MPI implementation, the locking mechanism involves
busy-waiting loops utilizing compare-and-swap, atomic fetch, and atomic
fetch-and-add operations to facilitate exclusive and shared locking. However,
this busy-waiting loop can potentially generate substantial network traffic in
typical Readers \& Writers  scenarios where the target memory is either
exclusively locked or intended to be exclusively locked, but shared locks are
already acquired.

Given these limitations, we propose two new DHT synchronization approaches.

\section{Improving Synchronization Methods for Distributed Hash Tables with MPI}

The following presents two new approaches to address the data consistency problem of the MPI-DHT:

\begin{itemize}
  \item A \emph{fine-grained locking} MPI-DHT which uses MPI's atomic
        operations.
  \item A \emph{lock-free} MPI-DHT with optimistic concurrency control.
        It uses \texttt{MPI\_Put} and \texttt{MPI\_Get} operations without
        any synchronization/locking. Instead, checksums are used to detect
        conflicts. This approach is adapted from Pilaf~\cite{2013-Pilaf}.
\end{itemize}

The decision to continue using the MPI library for the implementation
  of the DHT was guided by its status as the de facto standard in  HPC
  environments. This ensures the library's availability to a broad user base
  and provides a ``plug-and-play`` solution for every MPI-parallelized
application.

The two new MPI-DHT variants are discussed in detail in the following
sections.

\subsection{Fine-Grained Locking MPI-DHT}\label{secsec:fineDHT}

Regarding \emph{addressing} and \emph{collision} handling, the
coarse-grained locking (see Section~\ref{secsec:coarseDHT})
and the fine-grained locking MPI-DHT are identical.

  {\bf Data Consistency:} Instead of locking the entire memory window of
the target rank, the fine-grained locking mechanism locks only the
bucket that is currently to be accessed.  One way to approach this is
to define a memory window for each bucket.  This naive implementation
would lead to a massive overhead in terms of memory consumption, as
each memory window would return a data struct with a size dependent on
the MPI implementation. For Open MPI version 5.0.2, this would result
in a memory overhead of 352 bytes per bucket.  This is not feasible
for numerous buckets. Therefore, the implementation uses
self-implemented locking on 8 byte integers and relies on
\texttt{MPI\_Compare\_and\_swap} and \texttt{MPI\_Fetch\_and\_op}
operations. The implementation is adopted from the Open MPI
implementation of passive-target synchronization which uses UCX atomic
operations.\footnote{\url{https://github.com/open-mpi/ompi/blob/v5.0.6/ompi/mca/osc/ucx/osc_ucx_passive_target.c}}
In order not to violate the MPI standard by performing RMA operations
outside an epoch~\cite[p.~588]{mpi41}, all windows are {\em shared}
locked beforehand by all processes with
\texttt{MPI\_Win\_lock\_all}. This was inspired by the approach
employed by Schuchart et al.\ in their work on a lock-free message
queue~\cite{schuchartUsingMPI3RMA2019}.

A lock value of $0x10000000$ means that a writer is active on this bucket ({\em exclusive lock value}).
To acquire this exclusive lock, a write process attempts to set the
lock atomically to $0x10000000$ if the current value
is set to zero by employing the \texttt{MPI\_Compare\_and\_swap} function. This
is repeated until the lock is  acquired.

Readers only have to register their read interest.
Therefore,  a reader process attempts to
increment the lock atomically by one using the \texttt{MPI\_Fetch\_and\_op}
function until the lock is acquired. The acquisition of the lock is defined as
successful, when the value prior to the increment is less than $0x10000000$ (exclusive lock value).
However, if the value is greater than or equal to the exclusive lock value, the
origin process must revoke its read request by decreasing the lock value again and
attempting to acquire the lock once more. This guarantees that multiple readers
can access the same bucket concurrently, while only one writer can access the
bucket at a time.

The release of a lock is initiated by the origin process, which
performs a decrement on the lock value by the exclusive lock value
(writer) or by one (reader), respectively.

In order to facilitate the locking mechanism at the bucket level, an
8-byte lock is incorporated into each bucket, which is initialized to
0. Due to the requirement that locks be aligned to a multiple of 8
bytes, a padding of up to 7~bytes may be necessary.  This results in
a maximum overhead of 15~bytes per bucket in comparison to the
coarse-grained locking MPI-DHT.

\subsection{Lock-Free MPI-DHT}\label{secsec:lock-free}

As with the fine-grained locking mechanism, the lock-free MPI-DHT
implementation utilizes the same addressing and collision handling method as
previously defined. However, the locking mechanism is substituted by a checksum
calculation inspired by the work of Pilaf~\cite{2013-Pilaf}. Rather than adding
bytes for a lock, each bucket contains an additional 32-bit value for storing a
checksum.

  {\bf Data Consistency:} In case of a \texttt{DHT\_write} operation,
the origin process is responsible for calculating a checksum of the
key-value pair and appending it to the bucket data.

Subsequently, a reading process retrieves the bucket data and
recalculates the checksum.  If the recalculated checksum is equal to
the stored checksum in the bucket, the key-value pair is returned to
the application. In the event of a mismatch, the \texttt{MPI\_Get}
operation and checksum check is repeated.  If the checksums persist in
diverging, the bucket is flagged as {\em invalid} in the meta
information field.  In the event that a write operation is initiated
on an invalid bucket, the bucket can be overwritten with the new
key-value pair.

This checksum-based approach is a lock-free mechanism that ensures the
data consistency of the MPI-DHT. Consequently, it does not require
additional atomic operations or locks. As discussed
in Section~\ref{secsec:fineDHT}, all windows are locked by all processes with
\texttt{MPI\_Win\_lock\_all} prior to any RMA operation.

\section{Evaluation of the MPI-DHT with Different Synchronization Methods}

The next section describes our test bed. Then we present results with
synthetic benchmarks to investigate the different behavior for read and
write requests. This is done for two different benchmarks (uniform and
zipfian distribution) to test the influence of the distribution of
bucket accesses.  Finally, we present results where all three MPI-DHT
variants were tested as surrogate model within the reactive transport simulator POET~\cite{LKLLS21}.

\subsection{Test Bed}\label{sec:testbed}
All the benchmarks were conducted on the cluster of the Potsdam Institute for
Climate Impact Research
(PIK)\footnote{\url{https://www.pik-potsdam.de/en/institute/about/it-services/hpc}},
Germany. Each node is equipped with two AMD EPYC 9554 CPUs, each possessing 64
cores and a base clock speed of 3.1 GHz.
In all experiments, a dense mapping is used where all 128~cores of a node
are used.

Both CPUs share a total memory
capacity of 768 GB, with dual-ranked DDR5 memory being utilized. The nodes are
interconnected via NVIDIA Mellanox ConnectX-7 NDR Infiniband, which provides a
data transfer rate of 400 Gbps per port.

For the execution of all experiments, the GNU Compiler Collection in version
14.1 was employed. The Message Passing Interface (MPI) library of choice was
Open MPI version 5.0.6, accompanied by an UCX version of 1.17.0. The code was
compiled with the flags \texttt{-O3 -DNDEBUG}.

Furthermore, to enable Open MPI to utilize a single atomic remote memory
operation across the Infiniband network, the parameters
\texttt{osc\_ucx\_acc\_single\_intrinsic} and
\texttt{osc\_rdma\_acc\_single\_intrinsic} were set to true.

\subsection{Synthetic Benchmarks}

{\bf First experiment:}
The objective of the first benchmark is to evaluate the maximum
throughput of only read and write operations, respectively. To this
end, a benchmark was designed to generate a random number from which
an 80-byte key is derived.  The value size is set to 104~bytes. The
sizes were selected to model the size of a typical key-value pair in
the geochemical simulation POET~\cite{LKLLS21}. For random number
generation, both uniform and zipfian distributions were employed.
The zipfian distribution with  a skew of .99 and a range from 1 to 712,500
was chosen, since it models  best the distribution
of access requests within the POET simulation.

The benchmark generates 500,000~key-value pairs and writes them into
the DHT. After the completion of the write phase by all
benchmark processes, the same-key value pairs previously written are
read by each process. For both blocks, the throughput is calculated as
the number of operations per second.

  {\bf Second experiment:} The second benchmark is designed to evaluate
the performance of a mixed benchmark that is more read-intensive. For
this, a process performs 1,000,000 operations consisting of 95\% reads
and 5\% writes. The ratio also derives from the access pattern of the
geochemical simulation~\cite{LKLLS21}. The throughput is calculated as
the number of operations per second, combining both reads and writes.

\subsection{Results with Synthetic Benchmarks}

\begin{figure}
  \centering
  \subfloat[Read only]{\includegraphics[width=.5\linewidth,clip]{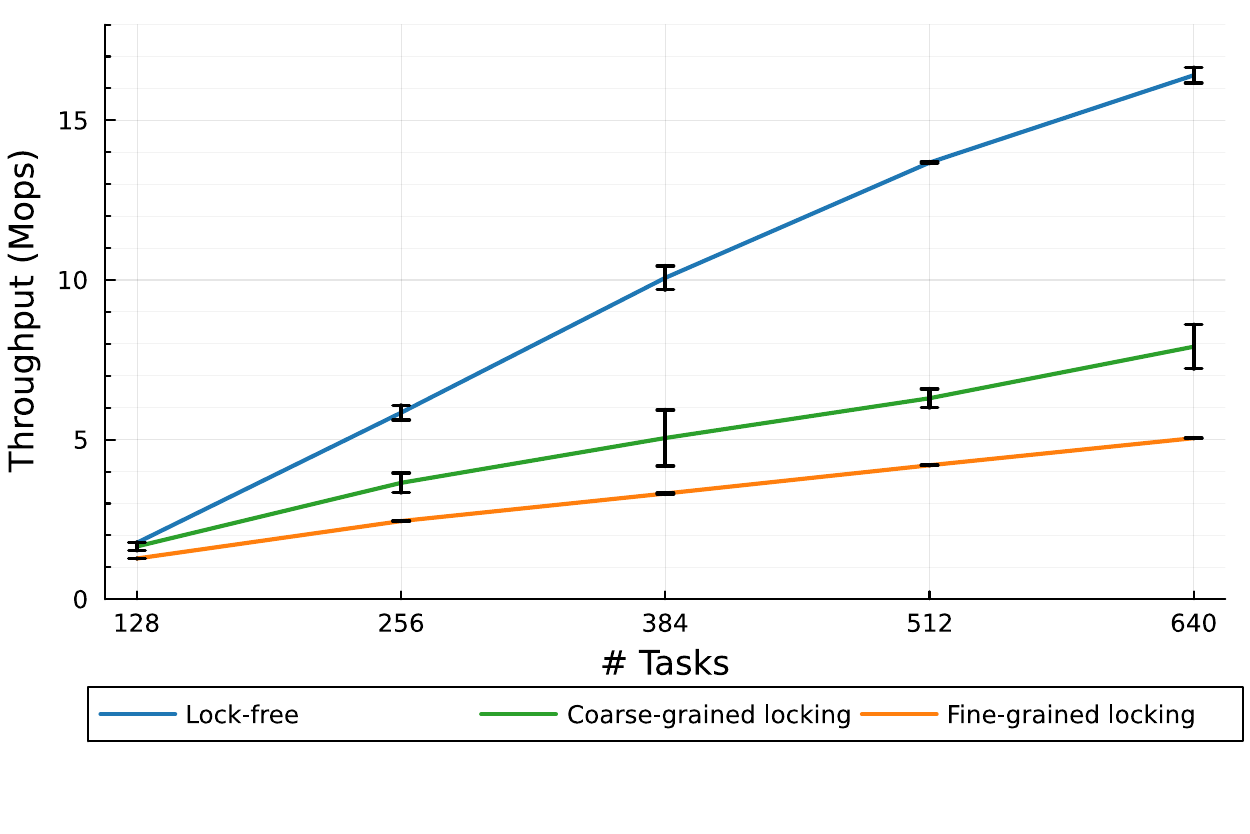}\label{fig:rw_read_uni}}%
  \subfloat[Write only]{\includegraphics[width=.5\linewidth,clip]{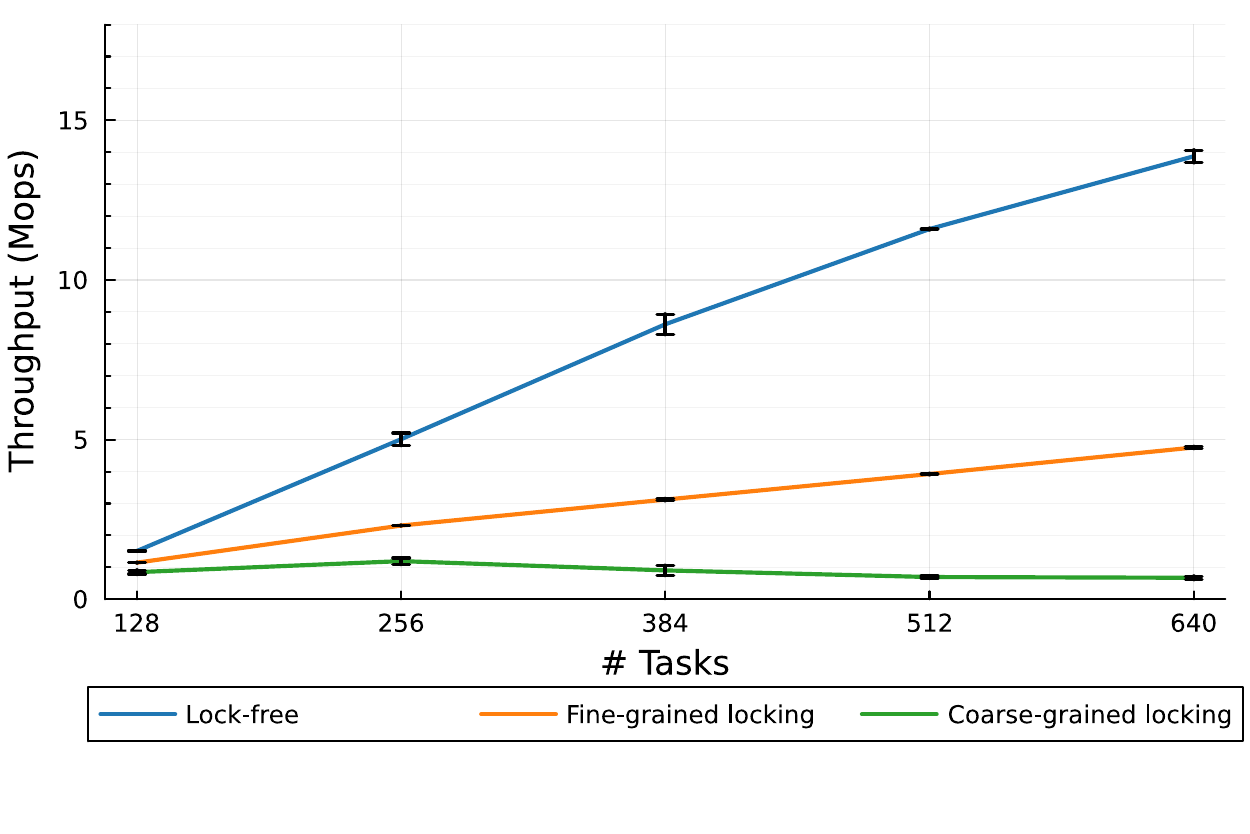}\label{fig:rw_write_uni}}%
  \caption{Throughput of read and write operations with \emph{uniform} distributed keys.}\label{fig:rw_uni}
\end{figure}

\begin{figure}
  \centering
  \subfloat[Read only]{\includegraphics[width=.5\linewidth,clip]{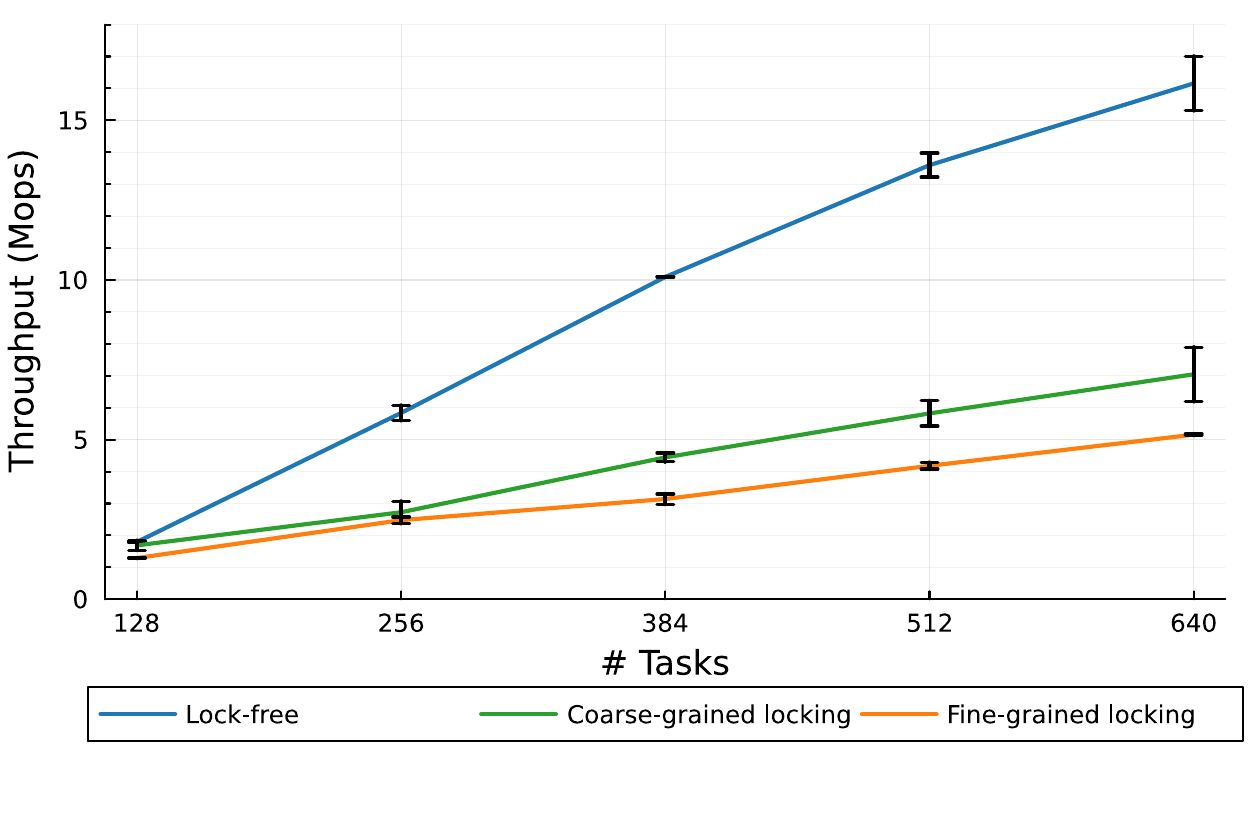}\label{fig:rw_read_zipf}}%
  \subfloat[Write only]{\includegraphics[width=.5\linewidth,clip]{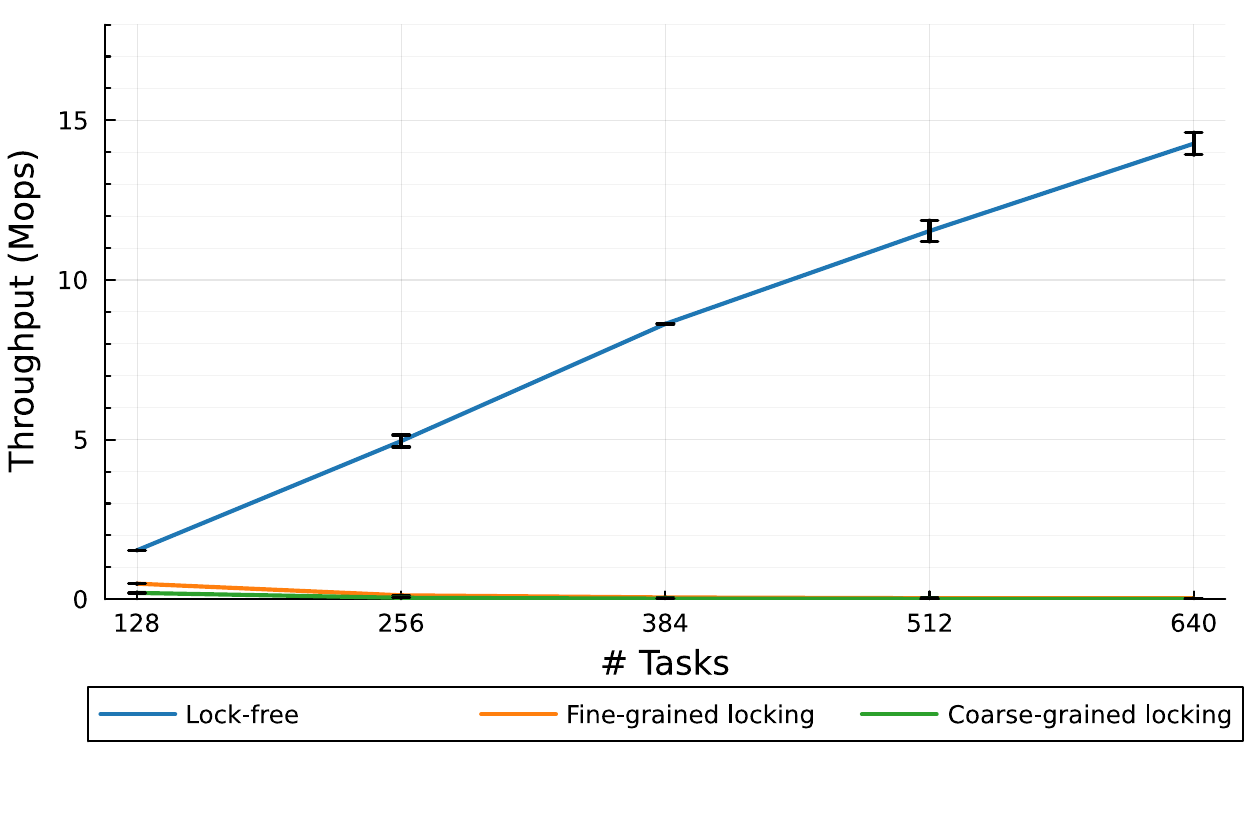}\label{fig:rw_write_zipf}}%
  \caption{Throughput of read and write operations with \emph{zipfian} distributed keys.}\label{fig:rw_zipf}
\end{figure}

We varied from 1 to 5 nodes with densely mapped NUMA nodes,
leading to a maximum of 640 processes. Each process provides 1 GB of memory for
the DHT. The benchmark was replicated five times, and the median values of the
throughput for read-only and write-only operations, respectively, along with
their respective standard deviations, are displayed in
Figures~\ref{fig:rw_read_uni} and~\ref{fig:rw_write_uni} for the uniform
distribution, and in Figures~\ref{fig:rw_read_zipf} and~\ref{fig:rw_write_zipf}
for the zipfian distribution.

\begin{figure}
  \centering
  \subfloat[Uniform Distribution]{\includegraphics[width=.5\linewidth,clip]{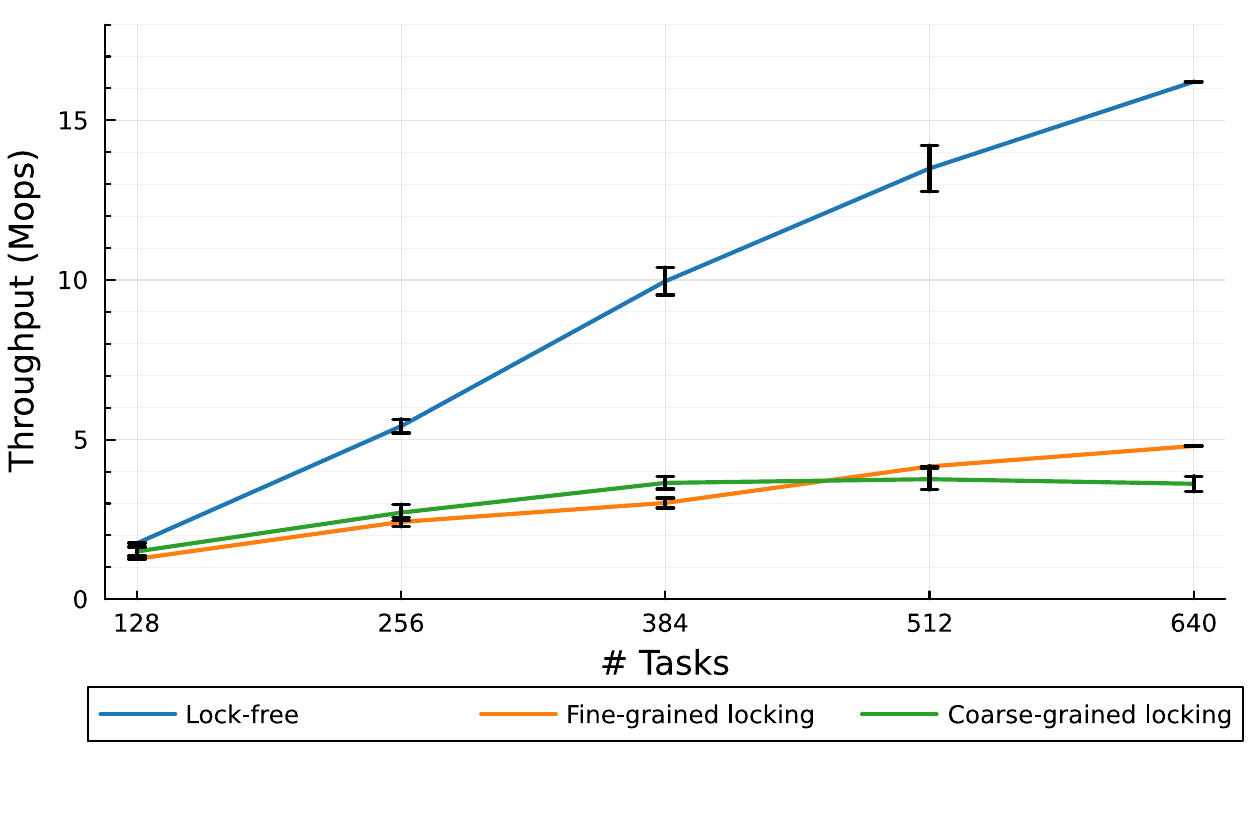}\label{fig:gs95_unif}}%
  \subfloat[Zipfian Distribution]{\includegraphics[width=.5\linewidth,clip]{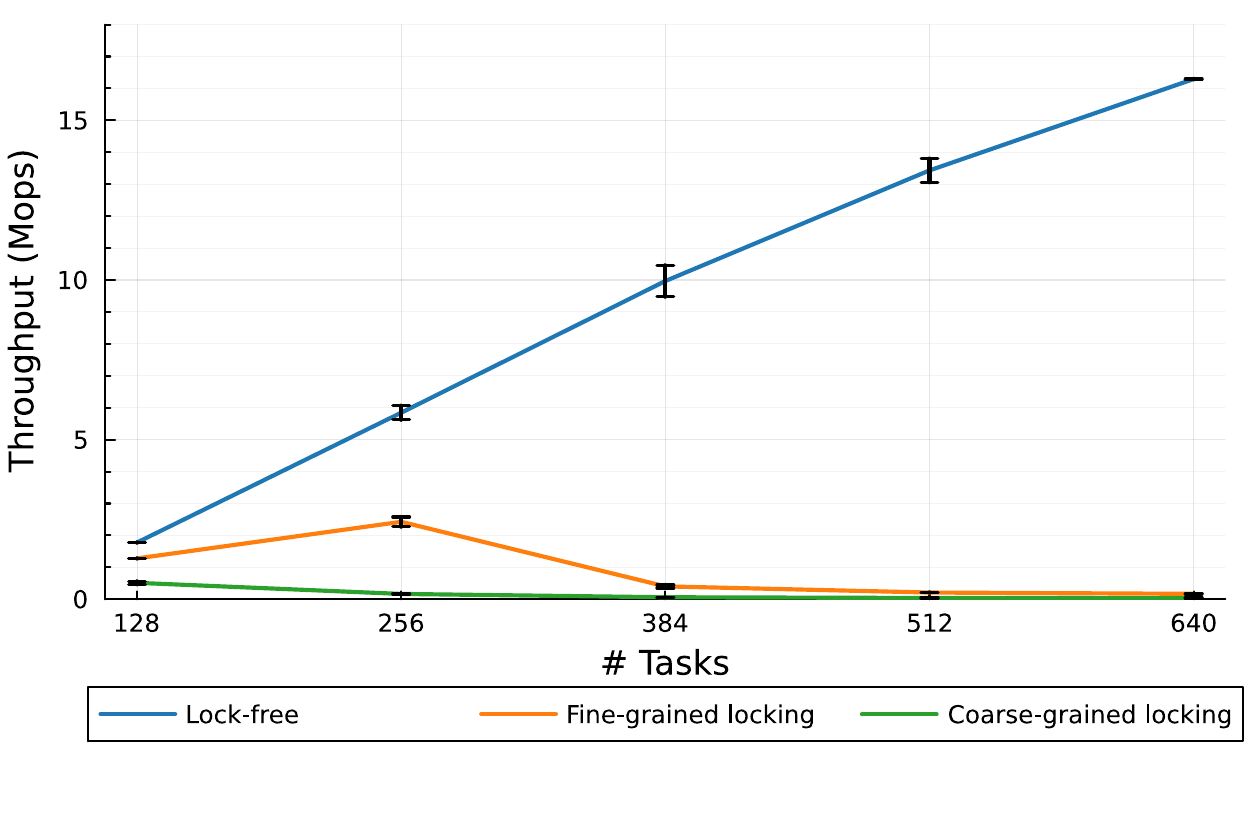}\label{fig:gs95_zipf}}%
  \caption{Throughput of mixed benchmark with a 95\% Read/5\% Write ratio
    for uniform and zipfian distributed keys.}\label{fig:mixed}
\end{figure}

For the read-only and write-only benchmarks, both uniform and zipfian
distributions, the throughput of the lock-free DHT is significantly higher
than both synchronized approaches. For example, in case of 640~processes
the lock-free MPI-DHT performs 16.4~(uniform distribution) resp.\ 16.2~(zipfian distribution) million \texttt{MPI\_READ} operations per second.
This is about 3~times higher than  fine-grained locking
and 2 times higher than coarse-grained locking.
Comparing read and write requests, the throughput of write operations is
always lower than that
of read operations for all approaches. This is expected since a write operation
requires at least one additional read operation to determine whether the bucket
is writable or matches the requested key to update the value.

\begin{table}[htb]
  \centering
  \caption{Write-only Performance for 640 Processes in Million Operations per Second.}\label{tab:write-only}
  \begin{tabularx}{\textwidth}{X c c c}
    \textbf{Benchmark} & \textbf{Coarse-Grained} & \textbf{Fine-Grained} & \textbf{Lock-Free} \\ \hline
    uniform            & 0.67                    & 4.75                  & 13.9               \\
    zipfian            & 0.01                    & 0.03                  & 14.3               \\
  \end{tabularx}
\end{table}

Table~\ref{tab:write-only} shows the write-only performance for
640~processes.  This demonstrates clearly the high overhead of the locking
mechanisms.  For uniformly distributed keys, the lock-free MPI-DHT
performs about a factor of 2.9~times better than fine-grained locking,
and about 20.6~times better than coarse-grained locking.  For the
zipfian distribution, the lock-free MPI-DHT is even 477~times better than
fine-grained and 1430~times better than coarse-grained locking.  Here,
the fine-grained locking mechanism shows it advantage of not blocking
other write operations destined for other buckets within the same
memory window.

For the mixed load benchmark, the throughput of the lock-free DHT is
in case of 640~processes 16.2~(uniform distribution) resp.\ 16.4~(zipfian distribution) which is
near its read-only performance.
For the uniform distribution, the performance
of the fine-grained locking  increases to
4.7 Mops compared to 3.6 Mops for coarse-grained locking.
Evidently, a zipfian distributed key challenges both synchronized approaches.
While fine-grained locking only increases from 1.27 to 2.42 million operations per
second for 128 to 256 processes, coarse-grained locking even degrades from 0.51 Mops
to 0.17 Mops for the zipfian distribution.

The number of reads that failed due to an invalid bucket caused by a checksum
mismatch as described in Section~\ref{secsec:lock-free} is shown in
Table~\ref{tab:artifical_checksum}. It can be seen that the read-only benchmark
does not produce any checksum mismatches, nor does the mixed load benchmark
with uniformly distributed keys. In fact, only zipfian's distributed key
generation results in reading corrupted data, caused by two or more processes
writing to a bucket at the same time. It also shows that the absolute number of
checksum mismatches increases as the number of processes increases. However,
these mismatches can be neglected because the number is very small in relative
terms.

\begin{table}[htb]
  \centering
  \caption{Checksum mismatches for the lock-free DHT.}\label{tab:artifical_checksum}
  \begin{tabularx}{\textwidth}{X c c c}
    \textbf{Benchmark} & \textbf{\# of Tasks} & \textbf{\# of Mismatches} & \textbf{Percentage [\%]} \\ \hline
    mixed - zipfian    & 128                  & 13                        & $1.1 \cdot 10^{-5}$      \\
    mixed - zipfian    & 256                  & 16                        & $6.5 \cdot 10^{-6}$      \\
    mixed - zipfian    & 384                  & 25                        & $6.8 \cdot 10^{-6}$      \\
    mixed - zipfian    & 512                  & 31                        & $6.3 \cdot 10^{-6}$      \\
    mixed - zipfian    & 640                  & 64                        & $1.1 \cdot 10^{-5}$      \\ \hline
    Others             & Any                  & 0                         & 0                        \\ \hline
  \end{tabularx}
\end{table}

\subsection{HPC Use-Case: POET}

POET~\cite{LKLLS21} is a coupled reactive transport simulator parallelized by MPI
that combines the flow and transport of solutes in porous media with
geochemical reactions. The version
used implemented an explicit upwind advection scheme as transport with constant
fluxes on a $500 \times 1500$ grid, homogeneous in species concentrations. The
same chemistry setup as described in~\cite{LKLLS21} was used. The simulation
was started with a constant injection of magnesium chloride by advection from
the top left boundary of the grid. As the magnesium concentration increases,
the previously equilibrated calcite begins to dissolve and dolomite
precipitates. Once calcite is completely consumed, dolomite also starts to
redissolve. POET reuses PHREEQC~\cite{parkhurstDescriptionInputExamples2013a}
to simulate these kinetic processes.

Due to the advective transport, a sharp reaction front can be observed during
the runtime of the simulation, while cells not yet reached by the reactive
solution remain unchanged. This makes it possible to cache previously simulated
results, e.g.\ in a distributed hash table. POET already implements caching with
the proposed DHT-API.

To cache simulation results, the input parameters for the geochemical
simulation are rounded to a user-defined number of significant digits to serve
as key for the DHT. These are 9 species and the simulation time step,
represented as double values. The stored value data consist of the
exact result simulated by PHREEQC together with the given input,
accounting for another 13 double values. Thus, the key-value pair has a
size of 80 bytes for the key and 104 bytes for the value.

The simulation was started with 500 time steps, where for each time step first
the flux (in this case constant fluxes), the transport (upwind advection) and
the geochemical reaction (PHREEQC) were simulated. There is actually one
chemical simulation per grid cell per time step. Whenever a cell is simulated
by PHREEQC, the result is stored in the DHT, and a next call with
approximately the same input to the time-consuming simulator can be avoided by
retrieving the value from the DHT.

\begin{figure}
  \centering
  \includegraphics[width=.7\linewidth,clip]{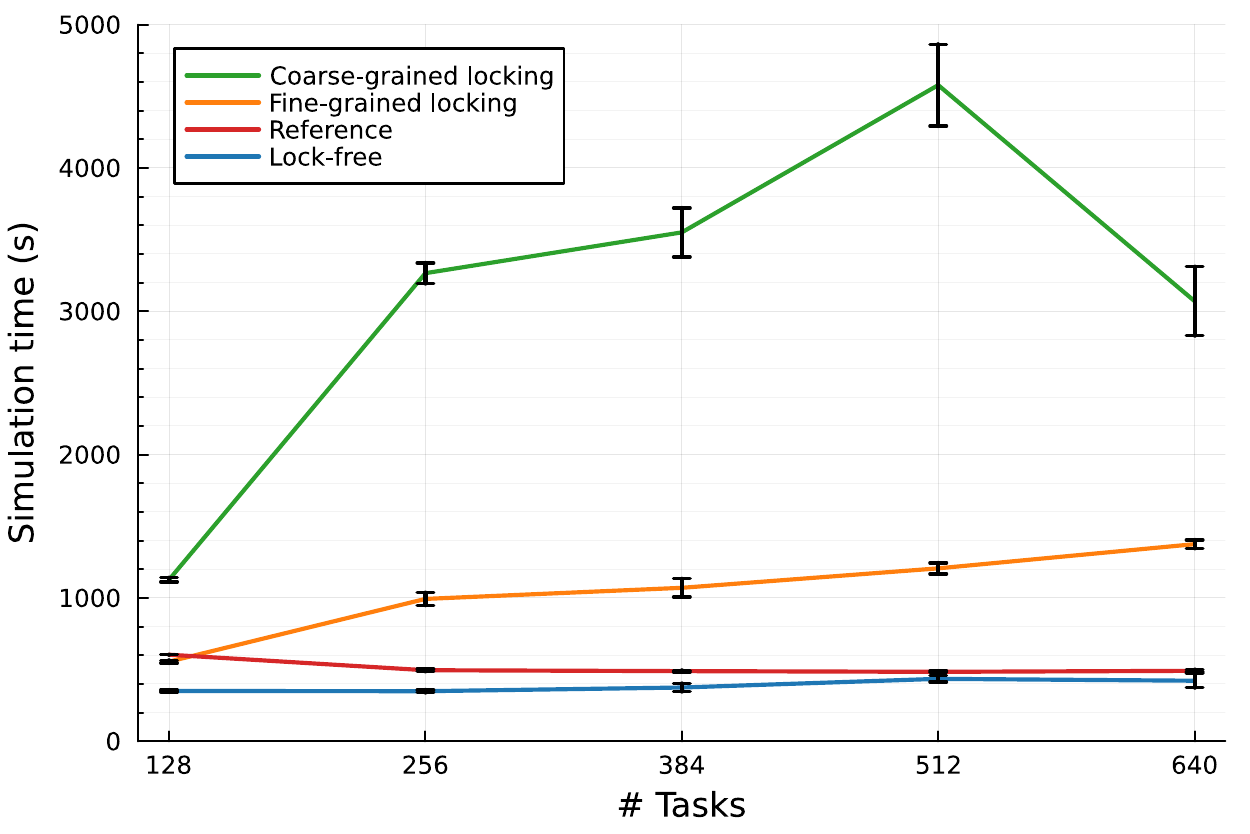}
  \caption{Runtime of the  chemical simulation of POET w/ and w/o DHT.}
  \label{fig:poet_simtime}
\end{figure}

Again, the simulation was scaled from 1 to 5 nodes, densely populated with one
MPI task per CPU core. Each experiment was repeated three times for the three
DHT approaches, and a reference without any DHT enabled. The median runtimes of
the chemical simulation, along with the respective standard deviation, are
shown in Figure~\ref{fig:poet_simtime}. For this particular simulation
setup, the average hit rate over all DHT runs was 91.8\%.

The coarse-grained locking DHT was a very good improvement in the
experiments presented in~\cite{LKLLS21}. The picture is now completely
different
on the newer hardware (see Section~\ref{sec:testbed}) and with the
newer faster POET version.
Now, only the lock-free DHT actually has a positive impact on
the runtime of the simulation.

\begin{table}[htb]
  \centering
  \caption{Performance gain of the POET simulation with lock-free MPI-DHT compared to the Reference Run without DHT.}\label{tab:poet_performance_gain}
  \begin{tabularx}{.45\textwidth}{c c}
    \textbf{\# of Tasks} & \textbf{Performance Gain} \\ \hline
    128                  & 41.9\%                    \\
    256                  & 29.5\%                    \\
    384                  & 23.3\%                    \\
    512                  & 10.1\%                    \\
    640                  & 14.1\%                    \\
  \end{tabularx}
\end{table}

The coarse-grained locking MPI-DHT slows down the application and shows
the worse runtime compared to the reference run for all node
counts. This approach also shows the highest variation between
repeated simulation runs.

The fine-grained locking MPI-DHT shows only a slight improvement
for 128 processes, while increasing the number of nodes
also degrades the performance compared to the reference run.

Furthermore, since increasing the number of processes has only a small impact
on the runtime of the reference run from 603 seconds for 128 processes to 491
seconds for 640 processes (speedup of 1.23 compared to the runtime on 128
cores), it can be assumed that the simulation has already reached the maximum
degree of parallelization when using only one node.  Nevertheless, the
lock-free DHT  is still able to reduce the runtime of the simulation.
The best result is achieved for 128 processes reducing the simulation
time from 603 seconds to 350 seconds compared to the reference
run. Table~\ref{tab:poet_performance_gain} shows the performance gain
for the different numbers of tasks.

Further, we investigated the number of mismatches which occurred during
the simulations with the lock-free DHT (see Table~\ref{tab:poet_checksum}).
The  maximum checksum mismatches are 4421 for 640~processes,
which means that only 0.0013\% of all read requests were corrupted.

\begin{table}[htb]
  \centering
  \caption{Checksum mismatches for POET simulation (lock-free MPI-DHT).}\label{tab:poet_checksum}
  \begin{tabularx}{.75\textwidth}{c c c}
    \textbf{\# of Tasks} & \textbf{\# of Mismatches} & \textbf{Percentage} [\%] \\ \hline
    128                  & 1507                      & $4.4 \cdot 10^{-4}$      \\
    256                  & 3049                      & $8.9 \cdot 10^{-4}$      \\
    384                  & 4315                      & $1.3 \cdot 10^{-3}$      \\
    512                  & 2884                      & $8.4 \cdot 10^{-4}$      \\
    640                  & 4421                      & $1.3 \cdot 10^{-3}$      \\
  \end{tabularx}
\end{table}

\section{Conclusion and Future Work}\label{sec:conclusion}

While server-based distributed key-value
stores are more common, the setup and maintenance costs of dedicated
hardware are questionable.  In a preliminary study, Intel's server-based
solution DAOS was outperformed by the distributed solution MPI-DHT with coarse-grained locking.
However, the MPI-DHT had scaling issues due to synchronization overhead.

Due to this fact, this paper investigates three MPI-Based DHT implementations with
different approaches to achieve data consistency: a coarse-grained locking DHT,
a fine-grained locking DHT, and a lock-free DHT with checksum validation.

First, an evaluation of these approaches was performed using a
synthetic benchmark.  The lock-free DHT  outperformed both locking
DHT approaches in terms of read and write performance,
with throughput improvements of up to 3 times for read operations and
up to 1,400 times for write operations.

The evaluation was extended to a real-world application where all
three approaches were integrated into the reactive transport simulator
POET as a surrogate model. Only the simulation runs with the lock-free
DHT showed runtime improvements compared to the simulation without
DHT. The runtime of POET was reduced by 14\% up to 42\% depending on
the number of processes.

The MPI-DHT does not support runtime table
resizing. However, resizing could be managed during HPC application
check pointing, adjusting the table size on restart.
In the experiments, only the Open MPI implementation was used. A
deeper analysis of the performance of different MPI implementations
will be considered in future work. Further, experiments with different
sizes of data values will be of interest.

In the experiments, the lock-free DHT impressed with a performance of
16 Million read operations per second and 15 Million write
operations per second.  This makes the lock-free DHT suitable as a fast data
cache within HPC applications.

The benchmark code and data are published on Zenodo~\cite{mpidht_data_2025} and the
MPI-DHT implementations are available as open-source software~\cite{mpidht_source_2025}.

\begin{credits}
  \subsubsection{\ackname}
  The authors gratefully acknowledge the
  Ministry of Research, Science and Culture (MWFK) of Land Brandenburg
  for supporting this project by providing resources on the high
  performance computer system at the Potsdam Institute for Climate Impact
  Research. Furthermore, the authors thank Klemens Kittan for his
  technical support.

\end{credits}

\bibliographystyle{splncs04}
\bibliography{references}

\end{document}